\documentstyle[aps]{revtex}

\begin{document}
\title{Popper's Experiment and Superluminal Communication}
\author{Edward Gerjuoy$^{*}$ and Andrew M. Sessler$^{**}$}
\address{$^{*}$ Department of Physics, University of Pittsburgh, Pittsburgh, PA.\\
15260; \\
$^{**}$ Lawrence Berkeley Laboratory, University of California, Berkeley, CA%
\\
94720.}
\date{July 11, 2005}
\maketitle

\begin{abstract}
We comment on Tabish Qureshi, ''Understanding Popper's experiment,'' AJP 
{\bf 73}, 541 (June, 2005), in particular on the implications of its Section
IV. We show, in the situation envisioned by Popper, that analysis solely
with conventional non-relativistic quantum mechanics suffices to exclude the
possibility of superluminal communication.
\end{abstract}

\section{INTRODUCTION.}

In a recent AJP paper$^{1}$ (hereinafter referred to simply as ``Q.'')
Tabish Qureshi has presented an analysis of an experiment proposed by Karl
Popper to test the standard interpretation of quantum theory. In this
Introduction we describe Popper's experiment and--because we believe
Qureshi's analysis could leave the readers of this journal with some
misconceptions--comment on Q. In Section II we show that in the situation
envisioned by Popper even conventional solely non-relativistic quantum
mechanics suffices to exclude the possibility of superluminal (faster than
light speed) communication. Some brief closing remarks are presented in
Section III.

Popper and Qureshi (see Q. Fig. 1) consider a source S emitting
non-interacting pairs of non-identical particles 1 and 2 moving
predominantly along the $x$ direction (horizontal), but with some small
components of momentum along the y direction (vertical) and with zero
components along the $z$ direction (perpendicular to $x$ and y); the
experimental situation pictured in Q. Fig. 1 is to be visualized as two
dimensional, therefore, lying in the $x,y$ plane only. The total momentum of
each pair is zero; also, any distribution in the components of momentum
along the $x$ direction is inconsequential, so that we are concerned solely
with the momenta $p_{1}=-p_{2}$ of particles 1,2 along the $y$ direction; We
assume, as Qureshi does in effect, that: (i) the source, at $x$ = 0, emits a
negligible number of particles with vertical momenta outside the range 
\mbox{$\vert$}%
$p_{1}|\leq P_{m}$; (ii) $P_{m}>0$ is much larger than any Heisenberg
Uncertainty Principle momenta realistically required to limit the spread of
the beam along $y$ in the region between A and B (see Q. Fig. 1); (iii) the
beam of particles 1, moving to the left, encounters at $x=-X$ a screen with
a narrow slit centered at $y=0$ (slit A of Q. Fig. 1); and (iv) this slit
introduces a momentum spread along $y$ that is much larger than $P_{m},$
with the result that after passing through slit A the particle 1 beam
spreads much more broadly along $y$ than it did before encountering slit A.
The question discussed by Popper and Qureshi is: Does conventional quantum
mechanics (what Qureshi calls the ''Copenhagen interpretation'') predict
that the particle 2 beam, moving to the right but not encountering a slit,
also will be spread much more broadly along $y$ at horizontal distances $x$ 
\mbox{$>$}%
$X,$ by virtue of the entanglement between particles 1 and 2 embodied in the
requirement that when emitted $p_{1}=-p_{2}?$

The unequivocal answer to this question, without the need to do any
calculating is ''No''. Indeed the observable effects of the beam on the
screen behind B (e.g., darkening as a function of $y)$ must in every respect
be completely independent of the size of the slit encountered at A.
Otherwise the observer at A (conventionally named Alice) could essentially
instantaneously transmit messages to her counterpart observer
(conventionally named Bob) viewing the screen behind B, now placed at a very
long distance $x$ 
\mbox{$>$}%
\mbox{$>$}%
$X$ ; in particular, if what Bob observes can depend on the size of the slit
Alice, using a code on which she and Bob had previously agreed, can send Bob
a message simply by widening and narrowing slit A. Such superluminal (faster
than light speed) communication of information is impossible.$^{2}$
Furthermore as Peres$^{3}$ has emphasized, conventional quantum mechanics
implies that it is impossible for Alice, by solely local operations, to
transmit any information whatsoever to Bob. Alice's control of the slit size
at A, without performing any operations whatsoever at any points between A
and the screen behind B, is a ''local operation'' by definition.

Unfortunately Section IV of Q. reasonably can be read to imply that Alice,
by detecting the passage of particles 1 through slit A as she controls the
width of the slit, can affect the spread of the beam on the screen behind B
(see in particular the text immediately following Eq. (13) of Q.). Qureshi
has assured us$^{4}$ that this reading is not his intention; rather, his
section IV is supposed to be concerned with coincidence measurements, on
particle 1 at slit A and on particle 2 at the screen behind B, performed on
a pair of particles that originally were simultaneously emitted from the
source S. In fact Qureshi, in an analysis$^{5}$ of Popper's experiment
written only a few months before Q. was submitted, explicitly states that in
the absence of such coincidence measurements the observable effects of the
particle 2 beam on the screen behind B will be independent of the width of
slit A. The clarifications, in this paragraph concerning the implications of
Section IV of Q., and in the preceding paragraph concerning the predictions
of conventional quantum mechanics respecting Popper's experiment, are among
our reasons for writing this paper.

\section{QUANTUM MECHANICAL ANALYSIS.}

We believe it will be useful to present a simple easily grasped (though
admittedly non-rigorous) derivation demonstrating that application of
conventional quantum mechanics to Popper's experiment predicts the
observable effects of the beam on the screen behind B must be completely
independent of the size of the slit encountered at A, or indeed of any other
local operations at A; thus, in Popper's experiment at least, local
operations cannot be employed to transmit information. No such derivation is
to be found in Qureshi's earlier paper,$^{5}$ nor in any other source of
which we are aware. It is important to note that manipulations of
measurement equipment at A, e.g., switching on electromagnetic fields in the
vicinity of A, are included in the ''local'' operations to which our
derivation pertains (see Subsections II.A and II.B), as are the performances
of actual measurements at A (see Subsection II.C). In our derivation,
however, actual measurement performances, which collapse the wave function
(Subsection II.C), require a different treatment than do all other local
operations, which affect the wave function via force terms generated in the
Schrodinger equation (Subsections II.A and II.B). Overall our derivation is
confined to local operations in Popper's experiment, although the analysis
in Subsection II.B does have wider application as will be seen. More general
proofs, not restricted to Popper's experiment, that information cannot be
transmitted by local operations can be found in the literature (cf., e.g.,
Peres$^{2}$ or Bruss$^{6}),$ but are difficult for non-experts.

It may seem surprising that a theorem of this nature can be established in
non-relativistic quantum mechanics. One might think that a proper
relativistic theory would be required to show the impossibility of
superluminal communication. What we show is that in Popper's experiment
''ordinary'' quantum mechanics precludes any local operations on particles 1
from changing any probability distributions in the entire beam of particles
2, no matter how entangled the particles are. This showing prevents
information about the manipulations of slit A, or about measurements
peformed by Alice, from being transmitted to Bob via the beams at any speed,
whether superluminally or relativistically allowed.

\subsection{Freely Moving Particles.}

For any given pair of particles 1 and 2 which simultaneously leave the
source S, the unnormalized wave function expressing the aforesaid
entanglement between them at the instant they leave the source is 
\begin{equation}
\Psi (y_{1},y_{2})=\int_{-\infty }^{\infty }dKW(K)e^{-iKy_{1}}e^{iKy_{2}},
\end{equation}
wherein: the plane waves have momenta $p_{2}=$ - $p_{1}=\hbar K;$ $W(K)$
describes the particle momentum distribution along the y direction; 
\mbox{$\vert$}%
$W(K)|^{2}$ is negligible for 
\mbox{$\vert$}%
$\hbar $K$|\geq P_{m};$ and the initial presumably random phase $e^{i\phi
(K)}$ multiplying each plane wave pair $e^{-iKy_{1}}e^{iKy_{2}}$ has been
absorbed in $W(K).$ Because every entangled particle pair moves
independently of every other such pair, the time evolution of this $\Psi
(y_{1},y_{2})$ predicts the probability distribution of all the particle 2
trajectories toward the screen behind B (see Q. Fig. 1) even though this $%
\Psi $ depends on the coordinates of only a single pair of particles.
Unhappily $\Psi $ given by Eq. (1) is not normalizable. Instead 
\begin{eqnarray}
\Psi ^{\dagger }\Psi &=&\int dy_{1}dy_{2}\int
dKW^{*}(K)e^{iKy_{1}}e^{-iKy_{2}}\int dK^{\prime }W(K^{\prime
})e^{-iK^{\prime }y_{1}}e^{iK^{\prime }y_{2}}  \nonumber \\
&=&(2\pi )^{2}\int dKdK^{\prime }W^{*}(K)W(K^{\prime })[\delta (K-K^{\prime
})]^{2}=(2\pi )^{2}\delta (0)\int dK|W(K)|^{2},
\end{eqnarray}
where: all the integrals in Eq. (2) [and all integrals below] run from -$%
\infty $ to $\infty ;$ herein and below the dagger $\dagger $ denotes the
adjoint; and, as is customary, $\delta (K)$ denotes the Dirac delta function
of $K.$ If $W(K)$ itself is normalized so that $\int dK|W(K)|^{2}=1,$ $%
|W(K)|^{2}dK$ legitimately can be interpreted as the probability that, when
the source emits a particle pair, the wave number of particle 2 will lie
between $K$ and $K+dK$ (still concentrating solely on motion along the $y$
direction, of course); henceforth we will assume that $W(K)$ has been so
normalized. The singularity on the right side of Eq. (2) can be avoided,
therewith making $\Psi $ normalizable, by confining the system vertically to
the region between the two distant horizontal planes $y=\pm L,$ at which
planes the momentum eigenfunctions in the expansion of $\Psi $ are required
to satisfy periodic (or other suitable) boundary conditions. This procedure
replaces the singular factor (2$\pi )^{2}\delta (0)$ in Eq. (2) by a well
behaved factor.

Introducing such boundary conditions, however, with the concomitant
requirement that integrals over all wave numbers be replaced by sums over
the allowed values of those wave numbers, leads to equations, e.g., the
analogs of Eqs. (1) and (2), which tend to obfuscate the transparency of our
analysis. We have decided not to impose boundary conditions, therefore,
believing it will be obvious that none of the inferences we draw from our
analysis are obviated either by our employment of the unnormalized $\Psi $
of Eq. (1) or by our retention of singular delta funtion factors as in Eq.
(2). One might think this unnormalizability can be simply dealt with by
appending a factor $\exp [-\gamma (y_{1}^{2}+y_{2}^{2})]$ to the right side
of Eq. (1), where $\gamma $ is a small positive constant. But the inclusion
of this factor means the initial wave function is not surely describing
particle pairs leaving the source S with equal and opposite momenta; for
such a source the dependence on $y_{1},y_{2}$ of the initial wave function
must be through the difference $y_{1}-y_{2}$ only, as is evident from Eq.
(1).$^{7}$ Once $W(K)$ has been normalized as described in the preceding
paragraph, $\Psi (y_{1},y_{2})$ can be formally ''normalized'' by appending
the singular factor [$(2\pi )\sqrt{\delta (0)}]^{-1}$ to the right side of
Eq. (1). We will denote this ''normalized'' $\Psi $ by $\Psi _{n},$ and will
employ the subscript $n$ to denote quantities calculated using $\Psi _{n}.$

The unit basis vectors $w(y,k)$ in wave number space (which by not requiring
the repeated inclusion of $\hbar $ is more convenient than momentum space),
satisfying $\int dkw^{*}(y,k)w(y^{\prime },k)$ = $\delta (y-y^{\prime }),$
are w$(y,k)=(1/\sqrt{2\pi })e^{iky}.$ Thus the components $\Phi
(k_{1},k_{2}) $ of $\Psi $ in wave number space are 
\begin{equation}
\Phi (k_{1},k_{2})=(1/2\pi )\int
dy_{1}dy_{2}e^{-ik_{1}y_{1}}e^{-ik_{2}y_{2}}\Psi (y_{1},y_{2})=2\pi \int
dKW(K)\delta (K+k_{1})\delta (K-k_{2})=2\pi W(k_{2})\delta (k_{1}+k_{2}).
\end{equation}
Then at the source the number of particles 2 with wave numbers between $%
k_{2} $ and k$_{2}+dk_{2}$ must be proportional to $D(k_{2})dk_{2},$ where
the particle 2 wave number distribution function 
\begin{equation}
D(k_{2})=\int dk_{1}|\Phi (k_{1},k_{2})|^{2}=(2\pi )^{2}\int
dk_{1}|W(k_{2})|^{2}[\delta (k_{1}+k_{2})]^{2}=(2\pi )^{2}\delta
(0)|W(k_{2})|^{2},
\end{equation}
a result whose total independence of our initial random phases $e^{i\phi
(K)} $ and proportionality to the probability $|W(k_{2})|^{2}$ supports the
validity of our analysis. Eqs. (2)-(4) imply that 
\begin{equation}
\Phi ^{\dagger }\Phi =\int dk_{1}dk_{2}|\Phi (k_{1},k_{2})|^{2}=\int
dk_{2}D(k_{2})=(2\pi )^{2}\delta (0)\int dk_{2}|W(k_{2})|^{2}=\Psi ^{\dagger
}\Psi ,
\end{equation}
as consistency of the analysis requires. It is readily seen that when $\Psi $
is replaced by $\Psi _{n}$ in Eq. (3), and the corresponding $\Phi _{n}$ is
employed in Eq. (4) to compute $D_{n},$ the factor $(2\pi )^{2}\delta (0)$
disappears from the right side of Eq. (4), yielding $D_{n}(k_{2})=$ $%
|W(k_{2})|^{2}.$ Thus $D_{n}(k_{2})dk_{2}$ is interpretable as the
probability at the source of finding particle 2 with wave number between $%
k_{2}$ and $k+dk_{2},$ consistent with the interpretation that ordinarily
would be afforded a calculation of the particle 2 wave number distribution
function starting from a properly normalized wave function. This additional
consistency additionally supports our belief, which we will not argue any
further, that the inferences we draw from our analysis in this Subsection
are valid although we use unnormalizable and singular functions.

Because the particles are assumed to move freely (without external
influences of any kind) in the space betwen A and B (see Q. Fig. 1), until
particles 1 reach A the number of their paired particles 2 with wave numbers
between $k_{2}$ and k$_{2}+dk_{2}$ surely continues to be proportional to $%
D(k_{2})dk_{2}.$ This obvious result can be derived formally by recognizing
that as long as particles 1 and 2 are moving freely their motions can be
thought to be described by the Schrodinger equation $i\hbar \partial \Psi
/\partial t={\bf \ }${\bf H}$\Psi ,$ whose formal solution when {\bf H} is
time-independent is$^{8}$ $\Psi (t)=e^{-i{\bf H}t/\hbar }\Psi (0).$ In the
present free particles 1,2 case {\bf H }is time-independent and{\bf \ = }$%
p_{1}^{2}/2m_{1}+$ $p_{2}^{2}/2m_{2}\equiv $ {\bf H}$_{1}${\bf \ + H}$_{2}.$
Then, since {\bf H} does not involve $K$ and {\bf H}$_{1},${\bf H}$_{2}$ act
respectively on $y_{1},y_{2}$ only, 
\begin{eqnarray}
\Psi (t) &\equiv &\Psi (y_{1},y_{2};t)=e^{-i{\bf H}t/\hbar }\Psi (0)=\int
dKW(K)[e^{-i{\bf H}_{1}t/\hbar }e^{-iKy_{1}}][e^{-i{\bf H}_{2}t/\hbar
}e^{iKy_{2}}]  \nonumber \\
&=&\int dKW(K)[e^{-i{\bf \hbar K}^{2}t/2m_{1}}e^{-iKy_{1}}][e^{-i{\bf \hbar K%
}^{2}t/2m_{2}}e^{iKy_{2}}]=\int dKW(K;t)e^{-iKy_{1}}e^{iKy_{2}},
\end{eqnarray}
where $\Psi (0)$ is $\Psi (y_{1},y_{2})$ from Eq. (1) and $W(K;t)=$ $e^{-i%
{\bf \hbar K}^{2}t/2\mu }W(K)$ with $\mu =m_{1}m_{2}/(m_{1}+m_{2}).$ The
right side of Eq. (6) has the same form as the right side of Eq. (1) except
that $W(K;t)$ has replaced $W(K);$ the components $\Phi (t)\equiv \Phi
(k_{1},k_{2};t)$ of the wave function in momentum space are defined as was $%
\Phi (0)\equiv \Phi (k_{1},k_{2})$ in Eq. (3) except that $\Psi (t)$
replaces $\Psi (0);$ and $D(k_{2};t),$ the particle 2 wave number
distribution function at time $t,$ is defined by Eq. (4) except that $\Phi
(t)$ replaces $\Phi (0).$ It follows that $D(k_{2};t)$ equals the right side
of Eq. (3), provided $W(k_{2};t)$ replaces $W(k_{2}).$ Therefore, since $
|W(k_{2};t)|^{2}=|W(k_{2})|^{2},$ it has been shown that $%
D(k_{2};t)=D(k_{2}),$ i.e., it has been shown (as just asserted) that as
long as the particles can be assumed to move freely the number of particles
2 with wave numbers between $k_{2}$ and k$_{2}+dk_{2}$ continues to be
proportional to $D(k_{2})dk_{2}.$ Moreover as Peres$^{9}$ has proved, when
the individual particles are represented by wave packets, as they should be,
the paired particles 1 and 2 emitted with opposite momenta do move in
opposite directions along the same straight line. Consequently, remembering
the paired particles are emitted with opposite momenta along $x$ as well as
along $y,$ until any burst of particles 1 reaches A the distribution
function $D(k_{2})$ of Eq. (4) determines the distribution--as a function of 
$y$--of any darkening or other observable effects produced by the
corresponding burst of particles 2 on any screen intercepting the particle 2
beam.

\subsection{Motion Under Forces.}

Now suppose that, as a consequence of some local operation at A other than
the performance of an actual measurement, the particles 1 no longer move
freely once they reach the vicinity of A. Then the Hamiltonian {\bf H }%
governing the particle motions still can be written in the form {\bf H = H}$%
_{1}+$ $p_{2}^{2}/2m_{2},$ but {\bf H}$_{1}$ (though of course still
independent of $y_{2}$ since the operation is local) now differs from $%
p_{1}^{2}/2m_{1}$ by terms depending on the particular local operation
adopted. If we justifiably are to visualize particles 1 as moving freely
until they reach the vicinity of A, these local operation terms should be
negligible unless $x$ is very close to the location $x=-X$ of A (as we have
been assuming to this juncture in this paper). But, once no longer
negligible, these local operation terms in {\bf H}$_{1}$ may be expected to
mix particle momenta along $x$ and $y,$ as well as to deflect particles 1
out of the $x,y$ plane in which we have assumed they move; certainly this is
what is likely to occur if the local operation involves electromagnetic
forces. It follows that for the purpose of determining the time dependence
of the particle motions when particles 1 are subject to local operations in
the vicinity of A, Eq. (1)--with its neglect of all particles 1,2
coordinates other than $y_{1}$ and $y_{2}$--generally is no longer useful.
Instead it is necessary to start from 
\begin{equation}
\Psi (0)\equiv \Psi ({\bf r}_{1},{\bf r}_{2})=\int d{\bf K}W({\bf K})e^{-i%
{\bf K\cdot r}_{1}}e^{i{\bf K\cdot r}_{2}},
\end{equation}
wherein $d{\bf K}$ = $dK_{x}dK_{y}dK_{z}$ and the notation should otherwise
be obvious.

We return to Eq. (7) below. For the moment, however, let us assume that the
local operation permits us to concentrate solely on motions along $y$ as we
have been doing, and as would be the case if the local operation were the
interruption of the particle 1 beam by a narrow horizontal slit at A. In
this event Eq. (6), giving the time dependence of the wave function in the
purely free particle case, legitimately can be replaced by 
\begin{equation}
\Psi (t)\equiv \Psi (y_{1},y_{2};t)=e^{-i{\bf H}t/\hbar }\Psi (0)=\int
dKW(K)u(y_{1},K;t)[e^{-i{\bf \hbar }K^{2}t/2m_{2}}e^{iKy_{2}}],
\end{equation}
where we are defining $u(y_{1},K;t)$ as the $y_{1}$ component of the
function $e^{-i{\bf H}_{1}t/\hbar }e^{-iKy_{1}},$ i.e., 
\begin{equation}
u(y_{1},K;t)\equiv [e^{-i{\bf H}_{1}t/\hbar }e^{-iKy_{1}}]_{y_{1}}=\int dy%
{\bf U}(y_{1},y;t)e^{-iKy},
\end{equation}
with {\bf U }the unitary operator $e^{-i{\bf H}_{1}t/\hbar }.$ Eqs. (8) and
(9) still assume the Hamiltonian operator {\bf H}$_{1}$ is time-independent.
When {\bf H}$_{1}$ is time-dependent, however, as for instance it would be
if Alice were to change the slit width at A while the beam of particles 1 is
impinging on A, it merely is necessary$^{8}$ to replace {\bf H}$_{1}t$ in $%
e^{-i{\bf H}_{1}t/\hbar }$ by the appropriately time ordered integral $%
\int_{0}^{t}dt^{\prime }{\bf H}_{1}(t^{\prime }).$ The key point is that the
right side of Eq. (9) remains a valid formula for $u(y_{1},K;t)$ in Eq. (8),
with {\bf U }still a unitary operator. Thus whatever the local operations,
time-independent or time-dependent, the components $\Phi (t)\equiv \Phi
(k_{1},k_{2},t)$ of the wave function in momentum space now are, recalling
Eq. (3) and using Eq. (9), 
\begin{equation}
\Phi (t)=(1/2\pi )\int dy_{1}dy_{2}e^{-ik_{1}y_{1}}e^{-ik_{2}y_{2}}\Psi
(t)=W(k_{2})^{-i{\bf \hbar }k_{2}^{2}t/2m_{2}}\int
dy_{1}e^{-ik_{1}y_{1}}u(y_{1},k_{2};t).
\end{equation}

Consequently, recalling Eq. (4), once particles 1 have reached A the number
of particles 2 with wave numbers between $k_{2}$ and k$_{2}+dk_{2}$ becomes
proportional to $D(k_{2};t)dk_{2},$ where the particle 2 wave number
distribution function now is 
\begin{eqnarray}
D(k_{2};t) &=&\int dk_{1}|\Phi (t)|^{2}=|W(k_{2})|^{2}\int dk_{1}\int
dy_{1}e^{-ik_{1}y_{1}}u(y_{1},k_{2};t)\int dy_{1}^{\prime
}e^{ik_{1}y_{1}^{\prime }}u^{*}(y_{1}^{\prime },k_{2};t)  \nonumber \\
&=&(2\pi )|W(k_{2})|^{2}\int dy_{1}\int dy_{1}^{\prime }\delta
(y_{1}-y_{1}^{\prime })u(y_{1},k_{2})u^{*}(y_{1}^{\prime },k_{2})=(2\pi
)|W(k_{2})|^{2}\int dy_{1}|u(y_{1},k_{2},t)|^{2}.
\end{eqnarray}
In Eq. (11), furthermore, remembering Eq. (9) and the fact that {\bf U} is
unitary, 
\begin{equation}
\int dy_{1}|u(y_{1},k_{2},t)|^{2}=u(t)^{\dagger }u(t)=[{\bf U}
e^{-ik_{2}y_{1}}]^{\dagger }{\bf U}e^{-ik_{2}y_{1}}=[e^{-ik_{2}y_{1}}]^{%
\dagger }{\bf U}^{\dagger }{\bf U}e^{-ik_{2}y_{1}}=\int
dy_{1}|e^{-ik_{2}y_{1}}|^{2}=(2\pi )\delta (0),
\end{equation}
where we have employed standard matrix manipulations. Eqs. (11) and (12),
taken together, make $D(k_{2},t)$ identical with $D(k_{2})$ from Eq. (4). It
again is readily seen that if the normalized $\Psi _{n}(0)$ replaces $\Psi
(0)$ in Eq. (8), the resultant $D_{n}(k_{2},t)$ = $%
|W(k_{2})|^{2}=D_{n}(k_{2}).$ In other words despite the local operations at
A that we have been discussing, the observable effects of the particle 2
beam on the screen behind B remain precisely what they would have been had
the particle 1 beam moved totally freely after leaving source S. In short,
such local operations do not permit Alice to send messages to Bob

The preceding key result that D$(k_{2},t)=D(k_{2})$ has been proved only for
those special local operations which do not affect the motions of particles
1 along the $x$ or $z$ directions; for more general local operations it is
necessary to start from Eq. (7) rather than Eq. (1), as we have explained.
But it now is readily seen that the above derivation of $D(k_{2},t)=D(k_{2})$
starting from Eq. (1) is directly parallelled by a derivation, starting from
Eq. (7), which--whether or not the paticles 1 move freely--yields $D({\bf k}
_{2},t)=D({\bf k}_{2}),$ where $D({\bf k}_{2})$ is the particle 2 wave
number distribution function at the source S for arbitrary wave number
vectors ${\bf k}_{2}$. Of course, when starting from the unnormalizable
three dimensional $\Psi ({\bf r}_{1},{\bf r}_{2})$ of Eq. (7), the equations
corresponding to Eqs. (2)-(6) and (8)-(12) contain three dimensional delta
functions rather than one dimensional delta functions. For instance the
equation corresponding to Eq. (4) is 
\begin{equation}
D({\bf k}_{2})=(2\pi )^{6}{\bf \delta }({\bf 0})|W({\bf k}_{2})|^{2},
\end{equation}
where $\delta ({\bf K})=\delta (K_{x})\delta (K_{y})\delta (K_{z})$ is the
three-dimensional Dirac delta function. Note in particular that $D({\bf k}
_{2})$ is proportional to $|W({\bf k}_{2})|^{2},$ just as $D(k_{2})$ is
proportional to $|W(k_{2})|^{2},$ so that regarding $D({\bf k}_{2})d{\bf k}$
as proportional to the number of particles 2 at the source with wave number
vectors lying in the range {\bf k}$_{2}$ to {\bf k}$_{2}+d${\bf k}$_{2}$ is
no less reasonable than was our corresponding interpretion of $%
D(k_{2})dk_{2}.$ As when starting from the one-dimensional Eq. (1), the
three-dimensional $\Psi $ of Eq. (7) can be made normalizable by confining
the system to the interior of the volume formed by the planes $x=\pm L,$ $%
y=\pm L,$ $z=\pm L$ and introducing suitable boundary conditions; in this
fashion the singular factor in Eq. (13) can be avoided, at the expense of
having to replace integrals over all wave numbers with sums over allowed
wave numbers only.

\subsection{Measurements by Alice.}

To this point, however, our derivation has not pertained to local operations
at A involving the performance of actual measurements. To see that
measurements at A also cannot enable Alice to send messages to Bob, let us
examine the consequences of a decision by Alice to make wave number
measurements of her own on the particle 1 beam, before Bob has a chance to
make his measurements. We want to see whether these measurements of Alice's
can alter our previous conclusions, e.g., our conclusion in Subsection II.A
that $D(k_{2})$ of Eq. (4) determines the distribution of wave numbers $%
k_{2} $ Bob observes irrespective of the local operations--of the sort,
e.g., considered in Subsection II.B--performed on particles 1. For this
purpose it is desirable to examine first an experimental situation which is
not complicated by the facts: (i) that $\Psi $ of Eqs. (1) or (3) is
unnormalizable; and (ii) that the unit basis vectors $w(y,k)$ in wave number
space [defined immediately preceding Eq. (3)] lie in the continuum. So
suppose that: (i) we again have entangled pairs of particles 1,2, with Alice
and Bob capable respectively of making local measurement observations on
particles 1 at A and on particles 2 at B; but (ii) at some instant the wave
function describing the state of a representative entangled pair 1,2 now is 
\begin{equation}
\Psi =\sum_{i,j}a_{ij}{\bf \alpha }_{i}{\bf \beta }_{j}.
\end{equation}
In Eq. (14): the ${\bf \alpha }_{i}$ are an orthonormal set of eigenstates
for the measurement operation Alice plans to make; the ${\bf \beta }_{j}$
are similarly defined for Bob; and $\Psi $ is normalized, implying the
numbers $a_{ij}$ satisfy $\sum_{i,j}|a_{ij}|^{2}=1.$ Then at this instant,
for any given $a_{ij},$ the quantity $|a_{ij}|^{2}$ customarily is regarded
as the probability that measurements on the particle pair will find particle
1 in the eigenstate ${\bf \alpha }_{i}$ and particle 2 in the eigenstate $%
{\bf \beta }_{j}.$ Correspondingly, summing $|a_{ij}|^{2}$ over all possible
states $i$ in which the particle 1 paired with this particle 2 might have
been found, one obtains the actual probability $\rho _{2j}$ of finding
particle 2 in the eigenstate ${\bf \beta }_{j},$ namely $\rho
_{2j}=\sum_{i}|a_{ij}|^{2}.$ If many independently moving entangled pairs
1,2 are being observed, the numbers of particles 2 in the various different
states ${\bf \beta }_{j}$ actually observed by Bob cannot but be
proportional to their respective probabilities $\rho _{2j}.$

The preceding paragraph has not specified whether or not Alice actually has
performed measurement observations on particle 1. Since nothing has been
said about any collapses of $\Psi $ induced by Alice's measurements, we
might infer that the preceding paragraph presumed Alice had not made any
actual measurements before Bob made his measurements. The important point,
which we are about to demonstrate, is that whether or not Alice did her
measuring before Bob is irrelevant to the validity of the above
interpretations of $|a_{ij}|^{2}$ and $\rho _{2j}.$ In particular suppose
Alice, before Bob makes any measurements on particle 2, observes that the
paired particle 1 is in the state ${\bf \alpha }_{i}.$ According to the
conventional understanding of measurements in quantum mechanics, this
measurement immediately collapses $\Psi $ of Eq. (14) into the new wave
function$^{10}$%
\begin{equation}
\Psi _{ci}={\bf \alpha }_{i}\{[\sum_{k}|a_{ik}|^{2}]^{-1/2}\}\sum_{j}a_{ij}%
{\bf \beta }_{j}.
\end{equation}
Evidently, except for the factor $[\sum_{k}|a_{ik}|^{2}]^{-1/2},$ $\Psi
_{ci} $ has plucked from $\Psi $ of Eq. (14) all the terms containing ${\bf %
\alpha }_{i}$ and only those terms, as one expects for the collapsed wave
function after observing particle 1 in the state ${\bf \alpha }_{i}.$ The
factor $[\sum_{k}|a_{ik}|^{2}]^{-1/2},$ which is consistent with the
so-called Born rule,$^{10}$ is required in order that $\Psi _{ci}$ be
normalized, i.e., in order that $\Psi _{ci}^{\dagger }\Psi _{ci}=1,$ as any
wave function describing an actual physical situation should be. According
to Eq. (14) the probability $\rho _{2j/i}$ of observing particle 2 in the
state ${\bf \beta } _{j},$ {\it knowing} that particle 1 has been observed
in the state ${\bf \alpha }_{i},$ is $\rho _{2j/i}$ = $|a_{ij}|^{2}[%
\sum_{k}|a_{ik}|^{2}]^{-1}.$ But, consistent with the preceding paragraph,
the probability $\rho _{1i}$ that Alice has observed particle 1 in the state 
${\bf \alpha }_{i}$ must be $\rho _{1i}$ = $\sum_{j}|a_{ij}|^{2}.$

Thus the probability of $Alice$ $first$ observing particle 1 in the state $%
{\bf \alpha }_{i}$ and{\it \ Bob only then} observing the paired particle 2
in the state ${\bf \beta }_{j}$ must be $\rho _{1i}\rho _{2j/i}=$ $%
|a_{ij}|^{2},$ exactly the probability, quoted in the penultimate paragraph,
for finding particle 1 in the eigenstate ${\bf \alpha }_{i}$ and the paired
particle 2 in the eigenstate ${\bf \beta }_{j}$ without any specified
temporal order in making the measurements on the two particles.
Correspondingly, since Alice had to find her particle 1 in some ${\bf \alpha 
}_{i},$ the actual probability that Bob will find the paired particle 2 in
the state ${\bf \beta }_{j}$ after Alice made her measurement again will be
the probability $\rho _{2j}=\sum_{i}|a_{ij}|^{2}$ obtained in the
penultimate paragraph. We conclude that when many independently moving
entangled pairs are being observed (as in Popper's experiment), the numbers
of particles 2 in the various different states ${\bf \beta }_{j}$ actually
observed by Bob will be proportional to the same respective probabilities $%
\rho _{2j}$ whether or not Bob has made his observations after measurements
by Alice. Note that this conclusion does not in any way depend on the nature
of the states ${\bf \alpha }_{i}$ and ${\bf \beta }_{j},$ i.e., does not
depend on the kinds of measurements Alice (on particles 1 only) and Bob (on
particles 2 only) have chosen to perform; it is assumed of course that the
measurements are performed independently, meaning that Bob receives no
communications from Alice which could enable him to modify his measurements
depending on Alice's measurement results. Therefore we have proved that when
the experimental situation involves many pairs of independently moving pairs
of entangled particles 1,2, and when the state of any representative
entangled pair is described by the wave function $\Psi $ of Eq. (14), Alice
cannot employ her local measurement observations on particles 1 at A to send
messages to Bob at B, because the nature of her measurements, and whether or
not she performs them, will not in any way alter Bob's observations of the
particles 2 at B.

The proof in the preceding paragraph, which we henceforth will term the
''foregoing'' proof, is quite generally valid for particle pair systems
described by Eq. (14), wherein $\Psi $ is normalized and is defined by a
discrete sum; for instance the foregoing proof is valid for the very
commonly discussed case of observations on a large number of similarly
entangled qubit pairs. Unfortunately, primarily because we have relied so
importantly on our ability to interpret the quantities $\rho
_{2j}=\sum_{i}|a_{ij}|^{2}$ arising from Eq. (14) as probabilities, we have
not been able to convincingly generalize the foregoing proof to this paper's
analysis of Potter's experiment, wherein the analogs of $\rho _{2j}$ are the
singular $D(k_{2})$ of Eq. (4) [in the one-dimensional case starting from
Eq. (1)] or the even more singular $D({\bf k}_{2})$ of Eq. (13) [in the full
three-dimensional case starting from Eq. (7)]; the facts that Eqs. (1) or
(7), and their respective succeeding Eqs. (4) or (13), involve integrals
rather than sums is a further complication. We argue as follows, however,
now confining our attention to the full three dimensional case starting from
Eq. (7): We already have pointed out that, via the device of confining the
system to the interior of the volume formed by the planes $x=\pm L,$ $y=\pm
L,$ $z=\pm L$ and imposing suitable boundary conditions, our analysis of
Potter's experiment could have been made to start with a wave function which
was normalized and involved a discrete sum, which wave function analog of $%
\Psi $ defined by Eq. (7) we will call $\Psi _{d}.$ The foregoing proof
unquestionably applies to this alternative formulation of Potter's
experiment starting with $\Psi _{d}.$ Moreover it is reasonable that with
arbitrarily large $L$ it should be possible to represent the actual physics
of a spatially confined experiment like Potter's with arbitrarily high
precision, even though the allowed particle wave numbers are limited to a
discrete set; certainly physicists have not hesitated to use discretized
wave expansions and box normalization ever since the dawn of quantum
mechanics.$^{11}$ To put it differently, since the allowed discrete wave
numbers are very close to each other for large $L$ and change as $L$
changes, it is unreasonable to think our rigorous proof--that Alice's
observations on particles 1 cannot affect Bob's observations on particles 2
when the source emits only allowed wave numbers for some particular
specified $L$--does not carry over to all wave numbers. We conclude that our
foregoing proof applies to Potter's experiment, whether Alice chooses to
make measurements on particles 1 when they still are moving freely (as in
Subsection II.A), or defers her measurements until particles 1 have reached
the slit A (as in Subsection II.B).

It is noteworthy that our derivation clearly implies any measurement
observations by Alice capable of collapsing the wave function may decrease
the range of {\bf k}$_{2}$ available to particles 2 on their way to the
screen behind B, but surely cannot increase this range.

\section{Concluding Remarks.}

The immediately preceding completes our demonstration that application of
conventional quantum mechanics to Popper's experiment predicts the
observable effects of the beam on the screen behind B must be completely
independent of the size of the slit encountered at A, or indeed of any other
local operations at A. We recognize that our derivation is not rigorous, but
believe it captures, in a fashion accessible to non-experts, the essence of
the physics involved in Popper's experiment when the particles involved are
not photons. We acknowledge that our demonstration is not convincing for a
Popper's experiment with pairs of photons, which do not obey the usual
Schrodinger equation and can be destroyed in the course of detection, a
possibility our derivation does not contemplate. That for photons it must be
possible to establish (though not necessarily to prove simply) theorems
similar to those derived in this paper follows from our general remarks in
the third paragraph of this paper.

Before closing we remark that the Schrodinger equation for freely moving
particles impinging upon a slit screen normally would be solved by imposing
some appropriate boundary condition at the screen (not at the slit of
course), in which event the equation $\Psi (t)=e^{-i{\bf H}t/\hbar }\Psi (0)$
will not yield the correct solution if {\bf H} is the usual free particle
Hamiltonian. This paper assumes the relevant physics of particles impinging
on a screen can be adequately reproduced via replacement of the boundary
condition with suitable forces; since such forces must exist, because
otherwise the particles simply would penetrate the screen, we do not doubt
the validity of our assumption, but believe we should make it explicit. Note
that merely postulating the existence of such forces describable by a
Hamiltonian is sufficient for our purpose; our proof in (Subsection II.B)
that $D({\bf k}_{2},t)=D({\bf k}_{2})$ depends only on the existence of an 
{\bf H }(whose details we need not know){\bf \ }which, when substituted for
the free particle Hamiltonian, will yield a $\Psi (t)=e^{-i{\bf H}t/\hbar
}\Psi (0)$ that correctly predicts the motion of the particles 1 in the
vicinity of the slit screen. Obviously a similar assumption must be made for
any other conceivable local operation at A that Alice might impose and which
at first sight is describable by a boundary condition, not by forces.

Finally we note that if $|W({\bf k}_{2})|^{2}$ in Eq. (13) is zero for k$%
_{2}=|$ {\bf k}$_{2}|$ 
\mbox{$>$}%
$P_{m}/\hbar ,$ no wave function collapsing measurements on particles 1, or
any other local operations on particles 1 for that matter, can result in any
particles 2 arriving at screen B along trajectories corresponding to k$_{2}>$
$P_{m}/\hbar .$ From this feature ALONE we can conclude that in Popper's
experiment inserting a narrow slit in the path of particles 1 will NOT cause
an increased spread in the angular trajectories of particles 2.

One of us (AMS) was supported by USDOE at the Lawrence Berkeley National
Laboratories Laboratory, under Contract No. DE-AC03-76SF00098.

\medskip

{\footnotesize 1. Tabish Qureshi, ''Understanding Popper's experiment.'' Am.
J. Phys. {\bf 73}, 541-544 (June, 2005).}

{\footnotesize 2. Asher Peres, ''Classical intervantions in quantum systems
II. Relativistic invariance,'' Phys. Rev. A {\bf 61}, 022117-1 to 8 (2000). }

{\footnotesize 3. Asher Peres, ''How the No-Cloning Theorem Got its Name,''
quant-ph/0205076, May 14 (2002). }

{\footnotesize 4. Tabish Qureshi, private communication, June 16, 2005.}

{\footnotesize 5. Tabish Qureshi, ''Popper's experiment. Copenhagen
Interpretation and Nonlocality,'' quant-ph/0301123, (March 30 (2004). This
paper is reference 18 of Q.}

{\footnotesize 6. D. Bruss {\it et al}., ''Approximate quantum cloning and
the impossibility of superluminal information transfer,'' Phys. Rev. A {\bf %
62}, 062302-1 to 4 (2000).}

{\footnotesize 7. Appending such a Gaussian factor is precisely how Qureshi
deals with the unnormalizability [see Q. Eq. (1)]. } {\footnotesize Hence it
is not evident that the results in his section IV pertain to actual
coincidence measurements in Popper's experiment.}

{\footnotesize 8. K. Gottfried, {\it Quantum Mechanics} (W. A. Benjamin, New
York, 1966), pp. 235-240. }

{\footnotesize 9. Asher Peres, ''Opposite momenta lead to oppposite
directions,'' Am. J. Phys. {\bf 68}, 991-992 (2000).}

{\footnotesize 10. N.D. Mermin, ''From cbits to qubits: teaching computer
scientists quantum mechanics,'' Am. J. Phys. {\bf 71}, 23-30 (2003), esp.
his Section VI discussion of the Born rule. See also the corresponding
discussion in E. Gerjuoy, ''Shor's factoring algorithm and modern
cryptography. an illustration of the capabilities inherent in quantum
computers,'' Am. J. Phys. {\bf 73}, 521-540 (2005), Section III.B.2. }

{\footnotesize 11. See, e.g., W. Heitler, {\it The Quantum Theory of
Radiation} (Oxford, 1954), pp. 38 ff.}

\end{document}